\documentclass[10pt,journal,final,twocolumn,letterpaper]{IEEEtran}

\usepackage{color}
\usepackage{amsmath,graphicx,xcolor,tabularx}
\usepackage{transparent}

\title{Time-Domain Multi-modal Bone/air Conducted Speech Enhancement}
\author{\IEEEauthorblockN{Cheng Yu,
Kuo-Hsuan Hung,
Syu-Siang Wang,
Yu Tsao,~\IEEEmembership{Member,~IEEE}, and \\
Jeih-weih Hung,~\IEEEmembership{Member,~IEEE}}
}

\begin{document}

\maketitle

\begin{abstract}
Previous studies have proven that integrating video signals, as a complementary modality, can facilitate improved performance for speech enhancement (SE). However, video clips usually contain large amounts of data and pose a high cost in terms of computational resources and thus may complicate the SE system. As an alternative source, a bone-conducted speech signal has a moderate data size while manifesting speech-phoneme structures, and thus complements its air-conducted counterpart. In this study, we propose a novel multi-modal SE structure in the time domain that leverages bone- and air-conducted signals. In addition, we examine two ensemble-learning-based strategies, early fusion (EF) and late fusion (LF), to integrate the two types of speech signals, and adopt a deep learning-based fully convolutional network to conduct the enhancement. The experiment results on the Mandarin corpus indicate that this newly presented multi-modal (integrating bone- and air-conducted signals) SE structure significantly outperforms the single-source SE counterparts (with a bone- or air-conducted signal only) in various speech evaluation metrics. In addition, the adoption of an LF strategy other than an EF in this novel SE multi-modal structure achieves better results.
\end{abstract}

\begin{IEEEkeywords}
multi-modal, bone/air-conducted signals, speech enhancement, fully convolutional network, fusion strategy
\end{IEEEkeywords}

\section{Introduction}\label{sec:intro}
Speech enhancement (SE) aims to improve the speech quality and the intelligibility in noisy environments and has been widely applied in numerous speech-related applications, such as automatic speech recognition \cite{Li2015Robust,dl-adv,chang2019mimo}, speaker recognition \cite{X-vec,speaker-v1, emo-rec1}, emotion recognition \cite{emo-rec1}, and assistive hearing devices \cite{Chen2016Large-scale, Lai2017Deep}, to improve the system robustness against environmental noises. Recently, deep learning-based models have been popularly used as a fundamental tool in SE systems \cite{chen2015speech,kolbk2017speech,xu2015regression,qi2020tensor,tan2019real,michelsanti2017conditional,araki2015exploring}. For most of these systems, the noise-corrupted speech is enhanced in the frequency domain using a deep-learning-based model with supervised learning. More specifically, the SE systems estimate the clean magnitude spectra or the corresponding ratio mask from the input noisy speech \cite{DDAE,sp-sep-prosody,sp-sep2,wang2018supervised,grais2018raw}. In addition to applying SE in the frequency domain, an fully-convolutional-network (FCN) SE approach \cite{pandey2019new,FCN2,gong2019dilated} has been proposed to directly estimate a time-domain waveform-mapping function to restore clean speech. A major advantage of this approach is the potential to circumvent the imperfect phase estimation issue. Additionally, the FCN-based SE system employs a smaller number of parameters than fully-connected-neural-network models, making it a suitable model to operate in mobile devices.

Recent studies have shown that SE systems can leverage complementary information, such as visual cues, as an auxiliary input to achieve better enhancement performance \cite{Gabbay2018,wu2016multi,chuang2019speaker,tawara2019multi}. As an alternative complementary information to visual cues, signal captured from a bone-conducted microphone (BCM) has the inherent capability to suppress air background noise to reduce the noise commonly recorded by an air-conducted microphone (ACM) \cite{shin2012survey,shimamura2006improving,lee2018bone,xing2018speech}. However, unlike an ACM-recorded speech signal, a BCM-captured waveform, in which the pronounced utterance is recorded through the vibrations from the speaker’s skull, may lose some high frequency components from the original spoken speech \cite{shin2012survey}. Several filtering-based and probabilistic solutions have been proposed to convert the BCM-recorded sound to its ACM version \cite{liu2018bone,tajiri2017noise,thang2006study}. The authors in \cite{Q-bone} proposed a reconstruction filter, which uses the long-term spectra of the speech, to perform the conversion. Meanwhile, some approaches have been proposed to combine ACM- and BCM-recorded sounds in the frequency domain with a linear transformation for SE tasks \cite{rahman2019multisensory,li2014multisensory}.

In this study, we propose a novel FCN-based SE method that leverages the acoustic characteristics of the signals recorded by a BCM in terms of the ensemble learning approach. In the proposed algorithm, the BCM-recorded waveform is used as an auxiliary input (with noise robustness property while low precision in the high frequency region), and combined with ACM to carry out SE in the time domain. Experiments were conducted on a personalized SE scenario and results show that the newly presented method yields significant improvements in terms of standardized objective metrics over the noisy baseline. These results clearly indicate that adequately integrating BCM- and ACM-recorded signals can help FCN models learn detailed harmonic speech structures, resulting in enhanced signals of high quality and intelligibility.

\section{Related works}\label{sec:part2}
We briefly review some novel studies that benefit a waveform-based SE task and/or exploit various signal sources.

\vspace{-0.2cm}\subsection{Deep learning-based model}
Employing a deep learning-based model structure is usually a main procedure of a state-of-the-art SE technique. In \cite{FCN2}, an FCN model was used to directly process the input time-domain waveform. By contrast, in the studies presented in \cite{TCNN, TDSE}, waveform-wise enhancement was conducted using a convolutional neural network (CNN) structure \cite{krizhevsky2012imagenet}. In comparison with a CNN, FCN only consists of convolutional layers, which can efficiently store information from the receptive fields of filters in each layer while possessing much fewer parameters. In addition, an FCN has been shown to outperform the conventional deep neural network (DNN), which consists of densely connected layers, for the application of SE.

\vspace{-0.2cm}\subsection{BCM/ACM conversion}
A straightforward method that can collect less distorted speech signals is to apply noise-resistant recording devices. As mentioned before, a BCM records signals through bone vibrations and is thus less sensitive to air background noise in comparison with an ACM. However, the BCM-recorded speech signals often suffer from a loss of high acoustic-frequency components, and this issue was addressed and partially alleviated through the BCM-to-ACM conversion technique applied in SE tasks \cite{Q-bone, AB-SE1,huang2017wearable}.

\vspace{-0.2cm}\subsection{Multi-modalality}
Another promising direction for waveform-based SE is to adopt a multi-modal system that extracts clean-speech information from various signal sources. In \cite{Stork1996SpeechreadingBH}, the authors proposed the use of audio-visual multi-modality in various speech-processing fields, and showed that integrating video modality with speech benefits various speech processing behaviors. The audio-visual system presented in \cite{hou2018audio} combines audio with lip-motion clips to access more bio-information and thereby promotes the SE performance. Despite the success of using audio-visual multi-modality for SE tasks, the corresponding high computational cost incurred and large amount of data storage required are obstacles for devices with limited computational resources. 

\section{proposed method}
\label{sec:part3}
In this section, we present a novel time-domain SE scenario that adopts multiple FCN models to fulfill the SE task. In particular, this novel scenario possesses multi-modal characteristics because it uses both BCM- and ACM-recorded signals. As is well known, the ACM-recorded signals contain complete (full acoustic-band) clean-speech information but are vulnerable to background noise, whereas the BCM-recorded signals possess a higher SNR but lack high acoustic-frequency components. Please note in this study, the ACM-recorded signal is the major source for the SE task while the BCM-recorded signal serves as the auxiliary source. Hence, we believe that, if arranged appropriately, the two types of signals can complement each other when applied to SE.

\vspace{-0.2cm}\subsection{The overall SE structure}
\begin{figure}[t!]
\begin{center}
\includegraphics[scale=0.8]{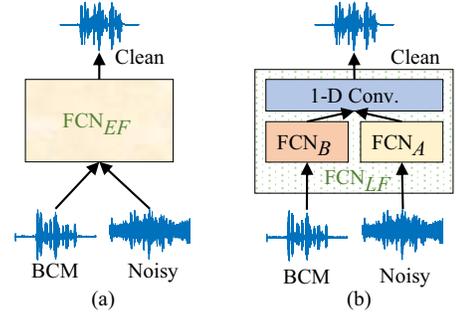}
\end{center}
{\vspace{-0.4cm}\normalsize\caption{Detailed structures of (a) EF strategy, FCN$_{EF}$, and (b) LF strategy, FCN$_{LF}$.\vspace{-0.2cm}}\label{fig:EF_LF}}

\end{figure}

A flowchart of the newly presented SE scenario is depicted in Fig. \ref{fig:EF_LF}, which indicates two different arrangements for the input BCM- and ACM-recorded signals. These two arrangements are created by either an early-fusion (EF) strategy or a late-fusion (LF) strategy. The difference between the EF and LF is in the stage during which the BCM- and ACM-wise representations are merged. In other words, the EF strategy suggests integrating BCM- and ACM-recorded raw waveforms at the very beginning of the SE framework to serve as the initial input, whereas in the LF strategy, the two signal sources are first individually processed, and the respective outputs are then brought together for a subsequent enhancement. To the best of our knowledge, determining which strategy is better for a multi-modal analysis mostly depends on the data types and tasks associated with the given multimedia dataset. In the following sections, we provide descriptions regarding the EF and LF arrangements shown in Fig. \ref{fig:EF_LF} in more detail.

\vspace{-0.2cm}\subsection{Early-fusion-strategy structure}
Following the EF strategy, the waveform-level BCM- and ACM-recorded noisy signals for each utterance in the training set are directly concatenated to form an input vector, which is used to train an FCN to approximate its noise-free ACM-recorded counterpart. The corresponding input-output relationship is therefore described as follows:
\begin{equation}
s_{EF} [n]= \mbox{FCN}_{EF}\{ x^{(A)}[n], x^{(B)}[n]\},
\end{equation}
where $x^{(A)}[n]$ and $x^{(B)}[n]$ with respect to the time index, $n$, represent the ACM- and BCM-recorded signals corresponding to an arbitrary noisy utterance; $\mbox{FCN}_{EF}\{.\}$ denotes the FCN model operation used, and $s_{EF}[n]$ is the enhanced signal expected to approximate the noise-free version of $x^{(A)}[n]$.

In this study, FCN$_{EF}$ is composed of seven hidden convolutional layers, each layer contains 30 kernels of size 55. In addition, to examine the impact of BCM, we construct another FCN model that is close to the FCN$_{EF}$ while it only adopts the ACM channel. Evaluations for these models will be described in Sec. \ref{sec:part4}. 

\vspace{-0.2cm}\subsection{Late-fusion-strategy structure}
In contrast to EF, the LF strategy first processes ACM- and BCM-recorded signals to obtain enhanced speech signals separately and then fuses the outputs from both sides. According to Fig. \ref{fig:EF_LF}(b), in the presented LF structure, we first create two FCN models to conduct a BCM-to-ACM conversion and an ACM-to-ACM enhancement, respectively, for noisy BCM- and ACM-recorded signals. The resulting output feature maps from both FCNs are then concatenated to serve as the input of another FCN model with simple 1-D convolutional layers, which are expected to produce nearly clean ACM-wise signals. The input-output relationship regarding the three FCNs in this LF multi-modal process can be expressed as follows:
\begin{gather}
s_{A}[n] = \mbox{FCN}_{A}\{ x^{(A)}[n]\},\\
s_{B}[n] = \mbox{FCN}_{B}\{ x^{(B)}[n]\},
\end{gather}
and
\begin{equation}
s_{LF}[n] = \mbox{FCN}_{LF}\{ s^{(A)}[n], s^{(B)}[n]\},
\end{equation}
where $\mbox{FCN}_{A}\{.\}$, $\mbox{FCN}_{B}\{.\}$ and $\mbox{FCN}_{LF}\{.\}$ denote the FCN operations for the ACM-to-ACM, BCM-to-ACM, and LF, respectively. In addition, $s_{A}[n]$, $s_{B}[n]$ and $s_{LF}[n]$
represent the output signals of the above three FCNs that share a common desired target, namely, a clean version of the ACM-recorded signal $x^{(A)}[n]$. The characteristics of each FCN model used here are further described as follows:

\begin{itemize}
\item The ACM-to-ACM enhancement FCN model, $\mbox{FCN}_{A}$, which aims to reduce noise distortions in the original ACM-recorded signals, is created following our recent study \cite{FCN2}. That is, the $\mbox{FCN}_{A}$ model has seven hidden layers, 33 filters for each layer and length 55 for each filter. According to \cite{FCN2}, this model can enhance ACM-recorded signals significantly.
\item Unlike $\mbox{FCN}_{A}$, the $\mbox{FCN}_{B}$ model conducting BCM-to-ACM conversion is designed in a compact manner, consisting of four convolutional layers in the order of (1, 257), (3, 1), (5, 15) and (1, 513), which follow the setting representation ``(number of kernels, kernel size)''.
\item The fusion function enhancing both $s_{A}[n]$ and $s_{B}[n]$ is constructed by two hidden convolutional layers with layer settings, (15, 55) and (1, 55), in the model.
\end{itemize}

\section{Experiments}
\label{sec:part4}
\subsection{Experimental setup}
\label{ssec:subhead}
We conducted the experiments on the Taiwan Mandarin hearing in noise test script (TMHINT) dataset \cite{TMHINT}. TMHINT is a balanced corpus consisting of 320 sentences and 10 Chinese characters in each sentence. The utterances in TMHINT were pronounced by a native Mandarin male speaker and recorded simultaneously with an ACM and a BCM in a silent meeting room at a sampling rate of 16 kHz. 

During the experiments, we split 320 utterances into three parts: 243 utterances for training, 27 utterances for validation, and 50 utterances for testing. In this study, we considered a personalized SE scenario, where only a small number of utterances pronounced by a single speaker was used for training and testing. For the training set, we added noise to the ACM-recorded utterances with several noise types (two talkers, piano music, a siren, and speech-spectrum-shaped (SSN) noise) at four SNR levels, -4, -1, 2 and 5 dB. For the test set, three noise types (car, baby-cry and helicopter), which were unseen noise types during the training, were added to ACM-recorded utterances at four SNR levels, -5, 0, 5 and 10 dB, to simulate mismatched conditions relative to the training set. Meanwhile, noise-free BCM signals were used as an assistant channel in both training and testing stages.

To evaluate the SE performance of the presented scenario, several objective metrics were used, including a perceptual evaluation of speech quality (PESQ) \cite{rix2001perceptual} with the wide-band configuration, short-time objective intelligibility (STOI) \cite{taal2011algorithm} and extended STOI (ESTOI) \cite{jensen2016algorithm}. PESQ indicates the speech quality with a score ranging from -0.5 to 4.5, whereas the STOI and ESTOI metrics typically having score range from 0 to 1 reflect the speech intelligibility (even though they might be also negative).
\begin{table}[!b]
\vspace{-0.4cm}\caption{Evaluation Scores of BCM signals and FCN$_{B}$.\vspace{-0.4cm}}\label{tab:fcnb}
\begin{small}
\setlength\tabcolsep{2pt}
\begin{center}
\begin{tabularx}{\columnwidth}{>{\centering}X|>{\centering}m{1cm}>{\centering}m{1cm}>{\centering}X|>{\centering}m{1cm}>{\centering}m{1cm}>{\centering\arraybackslash}m{1cm}}
\hline
\hline
& \multicolumn{3}{>{\centering}m{3cm}|}{\textbf{BCM}} & \multicolumn{3}{>{\centering}m{3cm}}{\textbf{FCN$_{B}$}} \\
\cline{2-7}
& \textbf{PESQ} & \textbf{STOI} & \textbf{ESTOI} & \textbf{PESQ} & \textbf{STOI} & \textbf{ESTOI} \\
\hline
\hline
\textbf{Avg.} & 1.247 & \textbf{0.619} & \textbf{0.395} & \textbf{1.554} & 0.608 & 0.362 \\ 
\hline
\hline
\end{tabularx}
\end{center}
\end{small}
\end{table}

\vspace{-0.2cm}\subsection{Evaluation results and discussions}
Several FCN-wise SE scenarios are compared here, including FCN$_{B}$ which applies a BCM-to-ACM conversion; FCN$_{A}$, which applies an ACM-to-ACM enhancement; and two novel multi-modal approaches, FCN$_{EF}$ and FCN$_{LF}$.

Table \ref{tab:fcnb} listed the metric scores for the original and the FCN$_{B}$-processed BCM-recorded utterances. From this table, we can see that the original BCM-recorded utterances exhibit a relatively low speech quality and intelligibility even though they do not encounter a noise distortion, which is primarily caused by a lack of high-frequency components. Next, the BCM-to-ACM conversion by the FCN$_{B}$ model moderately improves the speech quality in terms of PESQ scores, whereas the speech intelligibility is slightly worse than BCM.

\begin{figure}[!t]
\begin{center}
\includegraphics[width=0.9\columnwidth]{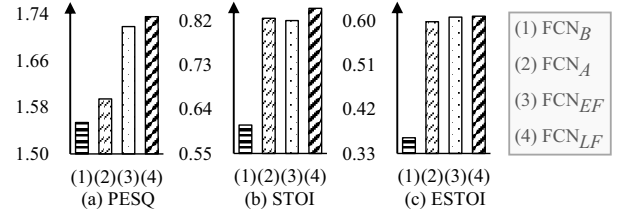}
\end{center}
\vspace{-0.3cm}{\normalsize\caption{Scores of different enhancement methods: FCN$_{B}$, FCN$_{A}$, FCN$_{EF}$, and FCN$_{LF}$ evaluated with (a) PESQ, (b) STOI, and (c) ESTOI.\vspace{-0.1cm}}\label{fig:bar}}
\end{figure}

\begin{table}[!b]
\setlength\tabcolsep{2pt}
\vspace{-0.4cm}\caption{Evaluation on noisy ACMs and FCN$_{A}$ in different SNR levels.\vspace{-0.4cm}}\label{tab:fcna}
\begin{small}
\begin{center}
\begin{tabularx}{\columnwidth}{>{\centering}X|>{\centering}m{1cm}>{\centering}m{1cm}>{\centering}X|>{\centering}m{1cm}>{\centering}m{1cm}>{\centering\arraybackslash}m{1cm}}
\hline
\hline
& \multicolumn{3}{>{\centering}m{3cm}|}{\textbf{Noisy ACM}}& \multicolumn{3}{>{\centering}m{3cm}}{\textbf{FCN$_{A}$}}\\
\cline{2-7}
& \textbf{PESQ} & \textbf{STOI} & \textbf{ESTOI} & \textbf{PESQ} & \textbf{STOI} & \textbf{ESTOI} \\
\hline
\textbf{10dB} & 1.722 & 0.912 & 0.750 & \textbf{1.965} & \textbf{0.915} & \textbf{0.761} \\
\hline
\textbf{5dB} & 1.452 & 0.849 & 0.624 & \textbf{1.682} & \textbf{0.877} & \textbf{0.673} \\
\hline
\textbf{0dB} & 1.273 & 0.766 & 0.500 & \textbf{1.446} & \textbf{0.809} & \textbf{0.552} \\
\hline
\textbf{-5dB} & 1.175 & 0.671 & 0.386 & \textbf{1.284} & \textbf{0.701} & \textbf{0.410} \\ 
\hline
\hline
\textbf{Avg.} & 1.405 & 0.799 & 0.565 & \textbf{1.594} & \textbf{0.826} & \textbf{0.599} \\ 
\hline
\hline
\end{tabularx}
\end{center}
\end{small}
\end{table}
\begin{table}[!b]
\vspace{-0.4cm}\caption{Evaluation on FCN$_{EF}$ and FCN$_{LF}$ in different SNR levels.\vspace{-0.4cm}}\label{tab:fcnfus}
\setlength\tabcolsep{3pt}
\begin{small}
\begin{center}
\begin{tabularx}{\columnwidth}{>{\centering}X|>{\centering}m{1cm}>{\centering}m{1cm}>{\centering}X|>{\centering}m{1cm}>{\centering}m{1cm}>{\centering\arraybackslash}m{1cm}}
\hline
\hline
& \multicolumn{3}{>{\centering}m{3cm}|}{\textbf{FCN$_{EF}$}} & \multicolumn{3}{>{\centering}m{3cm}}{\textbf{FCN$_{LF}$}} \\
\cline{2-7}
& \textbf{PESQ} & \textbf{STOI} & \textbf{ESTOI} & \textbf{PESQ} & \textbf{STOI} & \textbf{ESTOI} \\
\hline
\textbf{10dB} & 2.066 & 0.883 & 0.722 & \textbf{2.150} & \textbf{0.920} & \textbf{0.757} \\
\hline
\textbf{5dB} & 1.791 & 0.853 & 0.660 & \textbf{1.858} & \textbf{0.889} & \textbf{0.678} \\
\hline
\textbf{0dB} & \textbf{1.594} & 0.804 & \textbf{0.574} & 1.577 & \textbf{0.833} & 0.570 \\
\hline
\textbf{-5dB} & \textbf{1.422} & \textbf{0.744} & \textbf{0.475} & 1.357 & 0.740 & 0.433 \\ 
\hline
\hline
\textbf{Avg.} & 1.718 & 0.821 & 0.608 & \textbf{1.735} & \textbf{0.846} & \textbf{0.610} \\ 
\hline
\hline
\end{tabularx}
\end{center}
\end{small}
\end{table}

Next, the metric scores for the original noisy ACM-recorded utterances and their three enhanced versions (updated using FCN$_A$, FCN$_{EF}$ or FCN$_{LF}$) are listed in Tables \ref{tab:fcna} and \ref{tab:fcnfus}. From both tables, we can observe the following:
\begin{enumerate}
\item Comparing the results for ``BCM'' in Table \ref{tab:fcnb} and ``Noisy ACM'' in Table \ref{tab:fcna}, we see that BCM behaves worse than the unprocessed ACM noisy case, revealing that BCM alone fails to enhance speech signals.
\item The FCN$_A$ model, which was purely trained with ACM-recorded signals, behaves satisfactorily in promoting both quality and intelligibility of noisy ACM-recorded utterances. For example, improvements in the averaged PESQ, STOI, and ESTOI scores are $0.189$, $0.025$ and $0.034$, respectively.
\item Both multi-modal FCN$_{EF}$ and FCN$_{LF}$ structures, which integrate the information from both ACM and BCM, reveal higher PESQ, STOI, and ESTOI scores than the noisy baseline in all SNR cases. These results indicate the success of the presented multi-modal SE scenarios.
\item FCN$_{LF}$ achieves higher evaluation scores at high SNRs ($5$ dB and $10$ dB), and lower performances at low SNRs ($0$ dB and $-5$ dB) when compared with FCN$_{EF}$. One possible explanation is that FCN$_{EF}$ possesses better noise-robustness capability when it is employed in a more distorted situation.
\item FCN$_{EF}$ performs especially well and outperforms both FCN$_A$ and FCN$_{LF}$ for lower SNR cases ($0$ dB and $-5$ dB), but is less effective than FCN$_A$ in terms of STOI and ESTOI at SNRs of $5$ dB and $10$ dB. In comparison, FCN$_{LF}$ achieves better PESQ, STOI and ESTOI scores than FCN$_A$ under all SNR conditions.
\end{enumerate}

The evaluation scores from the previous tables averaged over different SNR cases are summarized in Fig. \ref{fig:bar} for comparisons. From this figure, we further confirmed that integrating speech sources from both BCM and ACM as in the FCN$_{EF}$ and FCN$_{LF}$ models, can achieve better SE performance in most noisy situations compared with FCN$_{A}$ and FCN$_{B}$, in which the models are created with a single speech source. Moreover, the LF strategy for multi-modal as in FCN$_{LF}$ appears to be a better choice here because it outperforms the others in all evaluation indices.

\begin{figure}[!t]
\begin{center}
\includegraphics[width=0.9\columnwidth]{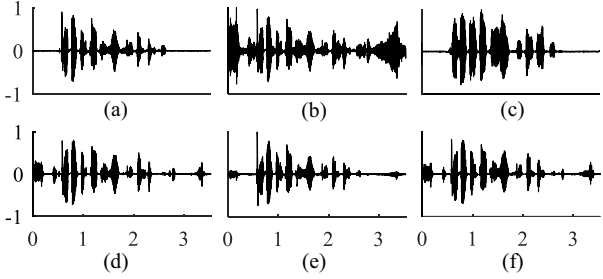}
\end{center}
\vspace{-0.3cm}\caption{The waveform of (a) clean ACM, (b) noisy ACM, (c) BCM, (d) noisy enhanced by FCN$_A$, and (e) FCN$_{EF}$ enhanced speech and the (f) FCN$_{LF}$ enhanced version.}
\label{fig:filter}
\end{figure}

\begin{figure}[!t]
\begin{center}
\includegraphics[scale=0.8]{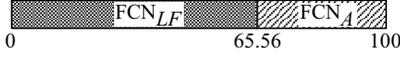}
\end{center}
\vspace{-0.3cm}\caption{The results (in percentage, \%) for the AB test that compares FCN$_{LF}$ and FCN$_{A}$ to determine which one brings less signal distortion.\vspace{-0.2cm}}
\label{fig:subj}
\end{figure}

In addition to the objective evaluations, subjective listening tests were also conducted to compare FCN$_{LF}$ and FCN$_{A}$. We selected two utterances from each of the 9 noisy conditions with 3 noise types (car, baby-cry and helicopter) at 3 SNR levels (-5, 0 and 5 dB), amounting to 18 ($2\times 3\times 3$) utterances. Each utterance was processed by either of FCN$_{LF}$ and FCN$_{A}$, thus generating 18 utterance pairs, which were used to perform single-blind-designed subjective AB tests in a quiet environment. 20 subjects with normal hearing were recruited to participate the tests. Each subject was asked to listen to a pair of processed utterances and select the one with lower signal distortion. The results are shown in Fig. \ref{fig:subj}, revealing that 65.56\% voted for FCN$_{LF}$ generating low signal distortions and the remaining 34.44\% voted for FCN$_{A}$ from the 360 tests (18 pairs $\times$ 20 subjects). In addition, we conducted a matched-pair T-test to confirm the improvement of FCN$_{LF}$ over FCN$_{A}$ as follows: First, we calculated the votes for FCN$_{LF}$ and FCN$_{A}$ for each of the 18 utterances from the 20 participants. Next, these 18 paired votes were used to determine the \textit{p}-value for the T-test, which equals $0.00088$. Such a small \textit{p}-value confirms that the improvement of FCN$_{LF}$ over FCN$_{A}$ is significant in the subjective evaluation tests.

Finally, Figs. \ref{fig:filter}(a)-(f) illustrate the waveforms of an utterance under six conditions: (a) clean ACM, (b) noisy ACM, (c) BCM-recorded clean, the noisy ACM enhanced by (d) FCN$_A$, and the concatenated BCM and noisy ACM signal enhanced by (e) FCN$_{EF}$ and (f) FCN$_{LF}$. When comparing the waveform of (c) with that of (a) in the figure, we can observe and confirm again that the BCM-captured speech is similar to the clean ACM counterpart while it has a smoother trajectory. Meanwhile, comparing Fig. \ref{fig:filter}(b) with Fig. \ref{fig:filter}(d), we see that FCN$_{A}$ can reduce noise distortion significantly. However, both FCN$_{EF}$ and FCN$_{LF}$ are shown to outperform FCN$_{A}$ by providing even less distorted signals when comparing Figs. \ref{fig:filter}(e)(f) and Fig. \ref{fig:filter}(d). As a result, we show that integrating BCM and ACM signals appropriately benefits a lot for SE as in the presented FCN$_{EF}$ and FCN$_{LF}$.

\section{Conclusion}\label{sec:part5}
In this study, we proposed a novel time-domain multi-modal bone/air conducted SE system, where BCM-signals were used as the auxiliary source to promote the conventional ACM-based SE system. We have derived and investigated two ensemble-learning-based fusion strategies, namely EF and LF, to perform multi-modal SE. In particular, for the LF multi-modal structure, two pre-trained FCN models (for BCM- and ACM-recorded signals, respectively) are concatenated and then followed by another compact FCN model with 1-D convolutional layers, along with the normalization and non-linear activation output layers. This structure provides signals with significantly improved PESQ, STOI, and ESTOI metric scores and consistently outperforms the FCN model which uses only ACM-recorded signals for training. Due to its compact model architecture as well as small input data size, the presented multi-modal scenario is quite suitable for implementations on mobile devices, such as cellphones, tablets, and even hearing aids. This study has revealed the effectiveness of the proposed multi-modal SE system on a speaker-specific scenario, such as mobile and personalized device applications. In the future, we will work to improve this system in its SE performance and extend its application in more severe environments.

\vfill\pagebreak

\bibliographystyle{ieeetr}
\bibliography{refs}

\end{document}